\documentclass[aps,twocolumn,preprintnumbers,amsmath,amssymb,nofootinbib,superscriptaddress,notitlepage]{revtex4}

\usepackage{graphicx,color}
\usepackage{relsize}
\usepackage{slashed}
\usepackage[normalem]{ulem}
\usepackage{tabu}
\usepackage{rotating}
\usepackage{bigstrut}
\usepackage{makecell}
\usepackage[dvipsnames]{xcolor}

\usepackage{amsmath}
\usepackage{amssymb,bm}
\usepackage{amsthm}
\usepackage{mathrsfs}
\usepackage{fancyhdr}
\usepackage{array}
\usepackage[all]{xy}
\usepackage{eufrak}
\usepackage{euscript}
\usepackage{enumerate}
\usepackage{slashed}
\usepackage{hyperref}
\usepackage{subfigure} 
\usepackage{mathtools}

\usepackage{soul}

\hypersetup{pdftex,colorlinks=true,linkcolor=blue,citecolor=blue,menucolor=black,urlcolor=blue,filecolor=blue}

\newcommand{\swu}{\affiliation{School of Physical Science and Technology,
	Southwest University, Chongqing 400715, China}}

\newcommand{\mev}{\rm{MeV}}

\begin{document}

\title{Low-energy $N\phi$ scattering from a pole-enhanced triangle diagram}

\author{Mao-Jun Yan}
\email{yanmj0789@swu.edu.cn}
\swu


\author{Chun-Sheng An}
\email{ancs@swu.edu.cn}
\swu

\author{Cheng-Rong Deng}
\email{crdeng@swu.edu.cn}
\swu

\date{\today}


\begin{abstract} 
We investigate low-energy $N\phi$ scattering driven by a pole-enhanced triangle-like diagram, in which the two-Kaon-exchange contribution is promoted by the near-threshold $\Lambda(1405)$ pole in the $N\bar K$ subsystem. 
Using an unphysical Kaon mass motivated by lattice simulations, we evaluate the $N\phi$ scattering length and find that this mechanism generates an attractive interaction with a magnitude of $-1.1$ to $-0.5\, \rm{fm}$. 
Spin-dependent effects are not treated explicitly and are expected to provide subleading corrections in the near-threshold region. 
We further analyze the low-energy behavior of the triangle-like diagram amplitude and show that the scattering length depends on the parameter $\delta$, defined as the mass difference between the $K\bar K$ threshold and the $\phi$ meson, and the pole position of $\Lambda(1405)$, where the $\Lambda(1405)$ plays a crucial role to understand $N\phi$ interaction.
Furthermore, by employing physical hadron masses, our calculated scattering length is found to be consistent with current experimental data, providing a unified description across both unphysical and physical mass regimes.
This type of interaction differs from that associated with van der Waals-type forces or the long-range tail of two-pion exchange, highlighting the role of three-body dynamics encoded in the pole-enhanced triangle-like diagram in shaping the near-threshold $N\phi$ interaction.
\end{abstract}

\maketitle

\section{Introduction}

OZI-suppressed interactions mediated by multi-gluon exchange have long been discussed in strong interactions.
In particular, such mechanisms were introduced to nucleon–heavy quarkonium scattering in terms of a QCD van der Waals–type interaction \cite{Hussein:1989ue,Brodsky:1989jd}, where an effective Yukawa potential,
$V_{(Q \bar{Q})}=-\alpha e^{-\mu r} / r$
was proposed to parameterize the interaction between a heavy quark–antiquark pair and the nucleon, with 
 $\mu$ and 
$\alpha$  characterizing the inverse range and strength.
Related attractive interactions have also been discussed in terms of soft-gluon exchange or two-pion exchange mechanisms, and have been explored in lattice QCD studies of nucleon–quarkonium systems \cite{Luke:1992tm,Brodsky:1997gh,Klingl:1998sr,Sibirtsev:2005ex,Fujii:1999xn,Wu:2024xwy,Yokokawa:2006td,Kawanai:2010ev,Alberti:2016dru,Lyu:2024ttm}.
These studies provide insights into possible OZI-suppressed interactions, whereas the present work focuses on a distinct threshold-driven mechanism relevant to the $N\phi$ system.

Motivated by earlier studies of nucleon-quarkonium interactions, analogous mechanisms have been explored for the $N\phi$ system by replacing the heavy quark pair with an $s\bar{s}$ configuration.
For this reason, a possible $N\phi$ bound state was first suggested in Ref.~\cite{Gao:2000az} by varying the parameters $\alpha$ and $\mu$ in the effective Yukawa potential $V_{(Q\bar{Q})}$.
Over the past two decades, the $N\phi$ interaction has been investigated using a variety of approaches, including quark models \cite{Huang:2005gw,Xie:2017mbe,An:2018vmk,Liu:2018nse}, coupled-channel analyses \cite{He:2018plt,Kim:2021adl,Sun:2022cxf}, lattice QCD simulations \cite{Lyu:2022imf}, and correlation-function analyses on both the theoretical and experimental sides \cite{Chizzali:2022pjd,Abreu:2024qqo,Kuroki:2024dtf,ALICE:2021cpv}.

Recent lattice QCD studies indicate that in the ${}^{4}S_{3/2}$ channel, inelastic transitions such as $N\phi \to \Sigma \pi K$ and $\Lambda \pi K$ are suppressed, and coupled-channel effects from $\Sigma K^{\ast}$ and $\Lambda K^{\ast}$ do not play a dominant role in determining the scattering length, yielding
$a = -1.43(23)_{\rm stat}\left({}^{+36}_{-06}\right)_{\rm syst}\,\mathrm{fm}$ \cite{Lyu:2022imf}.
An attractive interaction is also indicated by femtoscopic analyses from the ALICE collaboration, where a complex scattering length is extracted as
$a = -\left(0.85 \pm 0.34_{\rm stat} \pm 0.14_{\rm syst}\right)
- i\left(0.16 \pm 0.10_{\rm stat} \pm 0.09_{\rm syst}\right)\,\mathrm{fm}$ \cite{ALICE:2021cpv}.

Despite these advances, the underlying mechanism responsible for the $N\phi$
interaction remains model dependent and is still under debate
\cite{Chizzali:2022pjd,Kuros:2024dhc,Abreu:2024qqo,Feijoo:2024bvn,Kuroki:2024dtf}.
This motivates investigation of dynamical origins, in
particular those associated with $K\bar{K}$ threshold effect.

From a theoretical perspective, low-energy scattering processes are 
naturally described in terms of scattering amplitudes, where different
dynamics manifest themselves through their characteristic analytic
structures.
At the tree level, the $N\phi$ interaction is modeled by meson
exchange, involving pseudoscalar, scalar, and vector mesons.
For pseudoscalar-meson exchange, the contribution is expected to be suppressed
due to the small mixing angle between the $\eta$ and $\eta^{\prime}$ mesons.
In the case of scalar-meson exchange, although non-singlet $\sigma$ components
may couple to strange quarks~\cite{Peng:2020hql,Yan:2021tcp}, the singlet-octet
mixing in the scalar sector is known to deviate from the ideal limit that would
separate strange and non-strange components~\cite{Oller:2003vf}.
As a result, scalar exchange in the $N\phi$ channel is tiny due to the small SU(3) broken effect.
For vector-meson exchange, the small $\omega$-$\phi$ mixing angle,
$\epsilon_{\omega\phi}=0.0549$~\cite{Kucukarslan:2006wk}, further constrains the
possible strength of the interaction.
Moreover, when the spin-spin component is attributed to be  subleading relative to the
spin-independent part, the interaction is expected to be weak and
perturbative in the near-threshold region.



At the loop level, contributions beyond tree-level meson exchange may also appear. By analogy with the two-pion-exchange mechanism investigated in $J/\psi N$ scattering \cite{Fujii:1999xn,Wu:2024xwy}, two-pion exchange has been examined in $N\phi$ scattering and incorporated into the interaction potential within the HAL QCD \cite{Lyu:2022imf}, following the framework outlined in Ref.~\cite{TarrusCastella:2018php}.
The two-pion-exchange contribution is evaluated through
loop diagrams constructed from meson-baryon amplitudes.
However, the leading-order Weinberg-Tomozawa
interaction in chiral perturbation theory vanishes in $\phi\pi \to \phi\pi$
transition \cite{Roca:2005nm}.
Therefore, two-pion exchange in $N\phi$ scattering originates from subleading
chiral interactions and is expected to be suppressed in the low-energy
region.

Beyond the two-pion-exchange in $J/\psi N$ scattering, corrections from
open-charmed hadron loops, represented by triangle diagrams, are considered
in Ref.~\cite{Brodsky:1997gh}.
Given that, the coupling of the $J/\psi$ to $D\bar{D}$ is suppressed compared
to the $\phi K\bar{K}$ coupling, with a ratio $R_1 \simeq 1/4$.
In addition, the $DN$ interaction near threshold is weaker than the
$\pi N$ interaction, leading to a ratio $R_2 \simeq 0.1$.
In the strange sector, these suppression factors are naturalness.
The $\phi K\bar{K}$ coupling is unsuppressed,  corresponding to
$R_1 \sim \mathcal{O}(1)$, while the Weinberg-Tomozawa interaction enhances the
$N\bar K$ amplitude relative to $\pi N$ by a factor proportional to $m_K/m_\pi$.
Furthermore, the proximity of the $\phi$ mass to the $K\bar{K}$ threshold induces
a kinematical enhancement in the loop integral, which can be evaluated by a factor of
$\left(m_{J/\psi}-2m_D\right)^2/\left(m_\phi-2m_K\right)^2$
when compared to the corresponding $D$-meson loop.
These considerations indicate that triangle-diagram contributions associated with
the $K\bar K$ threshold are enhanced in the $N\phi$ system.
In particular, the proximity of the $K\bar K$ threshold, together with the presence
of near-threshold structures in the $N\bar K$ subsystem, promotes the loop
amplitude through both kinematics and dynamics.
As a consequence, the triangle diagram provides an effective interaction determined by $\sqrt{m_K(m_{\phi}-2m_K)}$ and $\Lambda(1405)$.
Such loop-induced mechanisms are expected to play a non-negligible role
in low-energy $N\phi$ scattering and contribute to the scattering
length within the framework of the effective-range expansion. In addition, a characteristic threshold scaling behavior as a smoking-gun signal of three-body dynamics is encoded in the pole-enhanced triangle diagram.


This article is organized as follows.
In Sec.~\ref{Sect2}, the estimation of the $N\phi$ scattering length
associated with triangle diagram contributions from two-Kaon exchange is introduced.
In Sec.~\ref{Sect3}, the behavior of the $N\phi$ scattering length is presented and
compared with results from HAL QCD and femtoscopic analyses by the ALICE collaboration.
A summary and outlook are given in Sec.~\ref{sect4}.

\section{formalism}\label{Sect2}

In the non-relativistic description of low-energy $N\phi$ scattering, the effective potential is introduced as
\begin{eqnarray}
V = c_0
+ d_0 \langle N | \vec{\sigma}\!\cdot\!\vec{k} | N \rangle
      \langle \phi | \vec{S}_1\!\cdot\!\vec{k} | \phi \rangle
= c_0 + \frac{1}{2}\, d_1^{\frac{1}{2},\,\frac{3}{2}}\,\vec{k}^{\,2},
\end{eqnarray} 
 which is an analogy to the heavy hadron interaction \cite{ Peng:2021hkr},
where $\vec{\sigma}$ and $\vec{S}_1$ denote the Pauli matrices and the spin-1 operator,
respectively, and $\vec{k}$ is the three-momentum of the nucleon in the center-of-mass
frame.
The $c_0$ represents the spin-independent (or spin-averaged) interaction,
while $d_1^{\frac{1}{2},\,\frac{3}{2}}$ parameterizes the spin-dependent contribution for
total spin $1/2$ and $3/2$, respectively.

In the vicinity of the $N\phi$ threshold, where $\vec{k}^{\,2}\to 0$, the inverse potential is expanded as
\begin{eqnarray}
\frac{1}{V}
&=& \frac{1}{c_0}
\frac{1}{1+\frac{1}{2}\frac{d_1^{\frac{1}{2},\,\frac{3}{2}}}{c_0}\vec{k}^{\,2}}
\simeq \frac{1}{c_0}
\left(1-\frac{1}{2}\frac{d_1^{\frac{1}{2},\,\frac{3}{2}}}{c_0}\vec{k}^{\,2}\right),
\end{eqnarray}
so that, at leading order in the threshold expansion, the spin-dependent correction
enters only at ${\cal O}(\vec{k}^{\,2})$ and is subleading.

The low-energy scattering amplitude is parameterized in
the effective-range expansion,
\begin{eqnarray}
\frac{8\pi\,th}{T} =\frac{1}{V+VGT} = -\frac{1}{a^{\frac{1}{2},\,\frac{3}{2}}} - i k ,
\label{eq:ERE}
\end{eqnarray}
where $th=m_N+m_\phi$, $V$ denotes the effective potential, and
$G=-\Lambda_0^{\frac{1}{2},\,\frac{3}{2}}/\pi + i k$ is the two-point loop function regularized by the
cutoff $\Lambda_0$ \cite{Dong:2020hxe}.
When $\Lambda_0^{\frac{1}{2},\,\frac{3}{2}}$ are truncated at the same scale, which is concerned here, $a^{\frac{1}{2},\,\frac{3}{2}}$ are degenerated to be $a$, where the contribution from the
spin-dependent interaction to the scattering length is attributed to be
subleading \cite{Dong:2021juy, Dong:2021bvy} and not considered in the present study.
To account for the off-shell configurations iterated through the two-point loop function $G$, the leading rescattering correction is evaluated. For a typical cutoff $\Lambda_0 \sim 200 \text{--} 500~\text{MeV}$, the scattering length is modified to
\begin{equation}
    -a_{N\phi} = V (1 + \mathcal{E}_{\text{off}}) ,
\end{equation}
where the theoretical uncertainty induced by the off-shell effects is bounded by $\mathcal{E}_{\text{off}} \equiv |V G V / V| \simeq 8\% \text{--} 23\%$. The smallness of this ratio indicates a rapid convergence of the iteration, ensuring that the effective potential $V$ evaluated on-shell robustly captures the dominant threshold dynamics.

\begin{figure}[ht]
\centering
\includegraphics[width=0.7\linewidth]{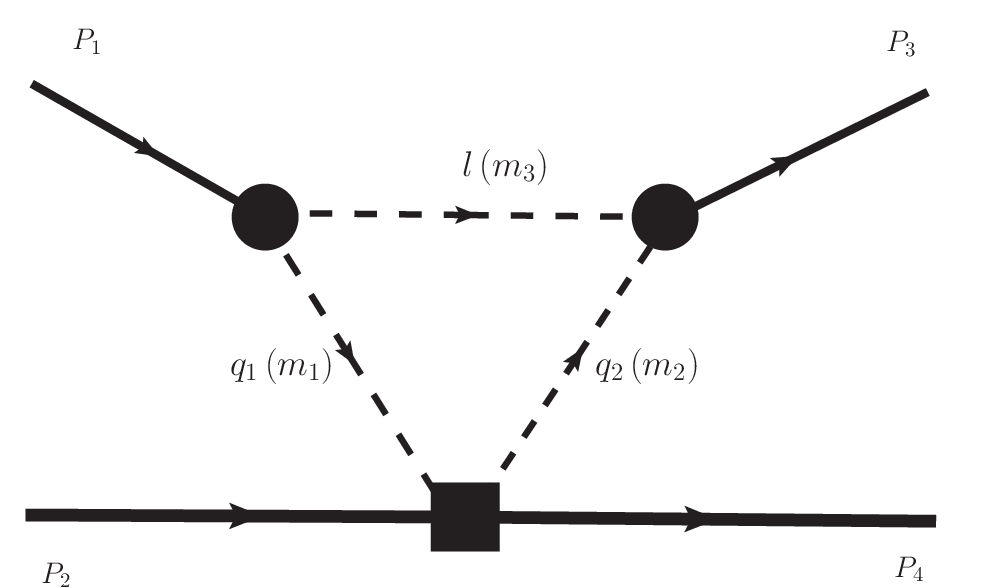}
\caption{The two-Kaon-exchange contribution to
$N\phi$ scattering through a triangle-like diagram.
The dashed, solid, and thick lines represent the Kaon, $\phi$ meson, and nucleon,
respectively. The corresponding particle mass and momentum are labeled in the figure.}
\label{fig:TriangleLoop1}
\end{figure}

The term $V$ corresponds to the two-Kaon-exchange contribution represented
by the triangle-like diagram shown in Fig.~\ref{fig:TriangleLoop1}.
This diagram involves two types of interaction vertices:
(i) the coupling of the $\phi$ meson to a $K\bar{K}$ pair, and
(ii) elastic $NK(\bar{K})$ scattering.
The former is described by a vector-pseudoscalar-pseudoscalar interaction,
with the effective Lagrangian given by
\begin{eqnarray}
\mathcal{L}_{\phi K\bar{K}} =
i g \, \phi^\mu \left(
\bar{K}\,\partial_\mu K - K\,\partial_\mu \bar{K}
\right),
\end{eqnarray}
where the coupling constant is fixed by
$g = m_V/(2F_\pi)=4.20$, with $m_V=m_\rho=775\,\mathrm{MeV}$ and
$F_\pi=92.1\,\mathrm{MeV}$.

\begin{figure}[ht]
	\centering
\includegraphics[width=0.950\linewidth]{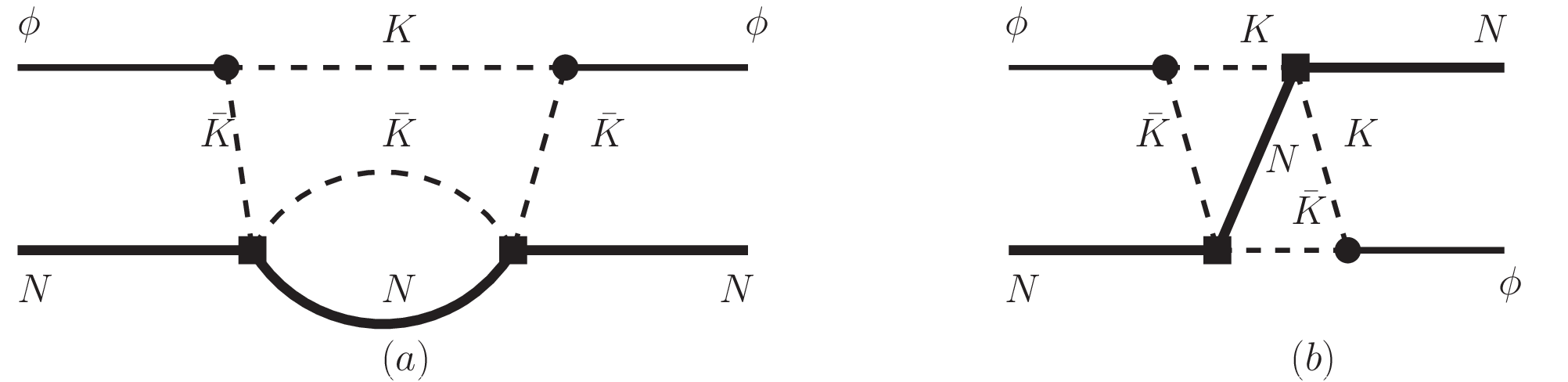}
	\caption{The $N\phi$ scattering driven by two-loop diagrams with Kaon scattering off nucleon. The dashed, solid, and bold lines represent Kaon, $\phi$ meson, and $N$, respectively, where the bubble represents $\mathcal{V}+\mathcal{V}\mathcal{G}\mathcal{V}\dots$ with the Weinberg-Tomozawa interaction $\mathcal{V}$ and $N\bar K$ loop $\mathcal{G}$.} 
	\label{fig:Twoloops}
\end{figure}

 The low-energy $NK$ and $N\bar{K}$ interactions entering the triangle diagram are
estimated in chiral perturbation theory, where poles are dynamically generated in $N\bar{K}$ scattering. In the isoscalar channel, the $\Lambda(1405)$ resonance is widely interpreted as a
dynamically generated $N\bar K$ bound state ~\cite{Lu:2022hwm,BaryonScatteringBaSc:2023zvt,BaryonScatteringBaSc:2023ori,
Guo:2023wes,Zhuang:2024udv,He:2024uau}.
The proximity of this pole to the $N\bar K$ threshold leads to a significant enhancement
of the low-energy $N\bar K$ amplitude, which in turn promotes the convergent part of the box-diagram contribution to $N\phi$ scattering.
In the isovector $N\bar K$ channel, a virtual pole has been reported in several analyses
and is sometimes associated with a broad $\Sigma$ excitation around
$1.4~\mathrm{GeV}$~\cite{Belle:2022ywa,Lu:2022hwm,Lyu:2024qgc,Wang:2024jyk,Li:2024tvo}.
This pole is located further from the physical sheet than the $\Lambda(1405)$ and
therefore provides a subleading correction to the box diagram at threshold.
By contrast, the $NK$ interaction in the $I=0$ and $I=1$ channels is governed by a
vanishing ($C_{ij}=0$) and a repulsive ($C_{ij}=2$) Weinberg-Tomozawa term, respectively,
and does not generate a near-threshold pole.
As a result, the $NK$ subsystem does not induce an analogous enhancement of the
convergent contribution.
These considerations indicate that the near-threshold structure
associated with the $\Lambda(1405)$ provides the leading enhancement mechanism in Fig. \ref{fig:TriangleLoop1}.
The pole of $\Lambda(1405)$ reads
\begin{eqnarray}
    t=-\lim _{s \rightarrow s_R} \frac{g_i^2}{s-s_R}, \label{eq:Lambda1405Pole}
\end{eqnarray}
with $N\bar{K}$ energy square $s$, pole position $\sqrt{s}_R=1417_{-4}^{+4}-i 24_{-4}^{+7}\,\rm{MeV}$ and  coupling $g_i=7.7_{-0.6}^{+1.2}\,\rm{GeV}$ \cite{Guo:2012vv}. These parameters bridge the Fig. \ref{fig:TriangleLoop1} and Fig. \ref{fig:Twoloops} (a).

\begin{figure}[ht]
\centering
\includegraphics[width=0.7\linewidth]{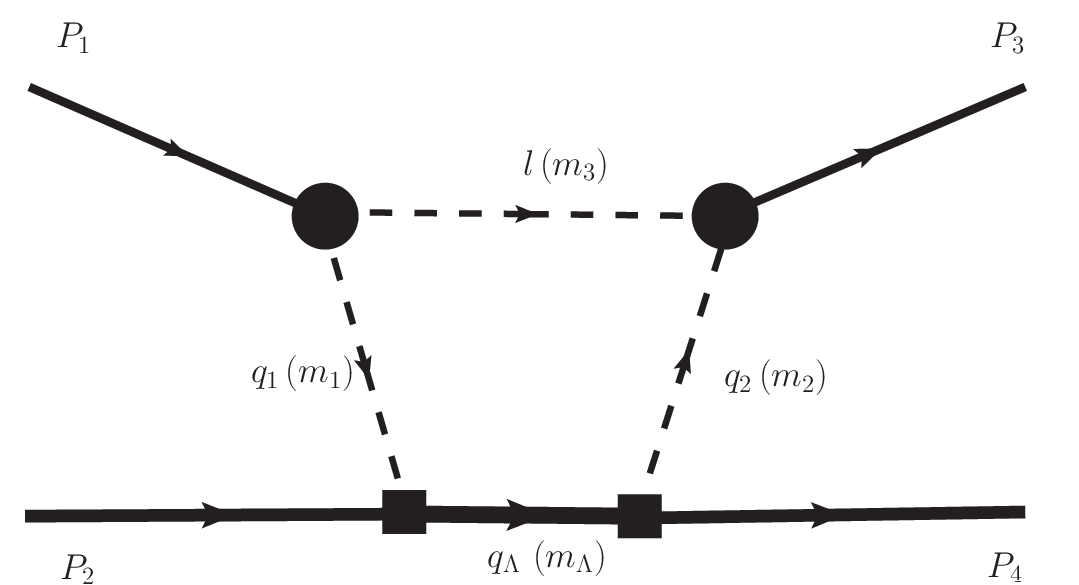}
\caption{Schematic illustration of the two-Kaon-exchange contribution to
$N\phi$ scattering through a box diagram.
The dashed, solid, thick, and thickened lines represent the Kaon, $\phi$ meson, nucleon, and $\Lambda(1405)$,
respectively. The corresponding particle mass and momentum are labeled in the figure.}
\label{fig:TriangleLoop}
\end{figure}

To evaluate the $N\phi$ scattering length, the full calculation involves evaluating the diagrams in Fig. \ref{fig:Twoloops}, which is very difficult, where the divergent parts in Fig. \ref{fig:Twoloops} (a) and (b) are regulator dependent.
However, the energy corrections on the particles in the loop in Fig. \ref{fig:TriangleLoop1} are small, and
the pole of $\Lambda(1405)$ in Eq. (\ref{eq:Lambda1405Pole}) is approached by a propagator of $\Lambda(1405)$ in the box diagram in Fig. \ref{fig:TriangleLoop}, where $s-s_R=2m_{\Lambda}\left(q^0_{\Lambda}-\omega_{\Lambda} \right)$.

Following the estimations of the vertices in the triangle-like diagram,
the numerator of the loop amplitude $I(l)$ associated with the box diagram in
Fig.~\ref{fig:TriangleLoop} is written as
\begin{eqnarray}
\mathrm{Num}
&=& -g_{\rm eff}\,\epsilon^\mu \left(2l-P_1\right)_\mu\left(2l-P_3\right)_\nu \epsilon^\nu\,u(P_2)\,\bar u(P_4) \nonumber\\
&=& -g_{\rm eff}\,
\left[4\,\vec{l}^{\,2}+\vec{P}_1\!\cdot\!\vec{P}_3\right]\,
u(P_2)\,\bar u(P_4),
\end{eqnarray}
with polarization vector $\epsilon^{\mu}$, 
where $g_{\rm eff}$ denotes the product of the coupling constants at the interaction
vertices.
In non-relativistic $N\phi$ scattering, the intermediate Kaons propagate close to their mass shell, such that
$l^0-m_K \sim \vec{l}^{\,2}/(2m_K) \ll m_K$.
The loop amplitude is decomposed into partial waves of the $N\phi$ system.
Near threshold, the S-wave component gives the leading contribution to the scattering
amplitude, while higher partial waves are suppressed by additional powers of the
external momenta.
In particular, the D-wave contribution enters at higher order in the threshold
expansion and does not contribute to the scattering length.

The S-wave projected amplitude $I(l)$ is 
\begin{widetext}
\begin{eqnarray}
I(l)
&=& \int \frac{d^4 l}{(2 \pi)^4}
\frac{-g_{\rm eff}\left[4\,\vec{l}^{\,2}+\vec{P}_1 \cdot \vec{P}_3\right]\,
u(P_2)\,\bar{u}(P_4)}
{\left(l^2-m_1^2\right)
\left[\left(P_1-l\right)^2-m_2^2\right]
\left[\left(P_3-l\right)^2-m_3^2\right] \left[ q^2_{\Lambda}-m_{\Lambda}^2 \right]} \nonumber\\
&=& \int \frac{d^4 l}{(2 \pi)^4}\,
\frac{-g_{\rm eff}}{8\,\omega_1\,\omega_2\,\omega_3}\,
\frac{4\,\vec{l}^{\,2}+\vec{P}_1 \cdot \vec{P}_3}
{l^0-\omega_1+i\epsilon}\,
\frac{u(P_2)\,\bar{u}(P_4)}{P_1^0-l^0-\omega_2+i\epsilon}\,
\frac{1}{P_3^0-l^0-\omega_3+i\epsilon} \frac{1}{2m_{\Lambda} \left(q^0_{\Lambda}-\omega_{\Lambda} + i\,\epsilon \right)}
,
\label{eq:Il}
\end{eqnarray}
\end{widetext}
where $m_1=m_2=m_3=m_K$ and the energies of the intermediate Kaons and $\Lambda(1405)$ are approximated as
\begin{eqnarray}
\omega_1 &=& m_K + \frac{\vec{l}^{\,2}}{2m_K}, \,
\omega_2 = m_K + \frac{(\vec{P}_1-\vec{l})^{2}}{2m_K}, \nonumber\\
\omega_3 &=& m_K + \frac{(\vec{P}_3-\vec{l})^{2}}{2m_K},\, \omega_{\Lambda}=m_{\Lambda}+ \frac{\left(\vec{P}_1+\vec{P}_2-\vec{l} \right)^2}{2 m_{\Lambda}}
\end{eqnarray}
which implies the integration over the component $l^0$ is convergent, and can be evaluated by contour integration.

Evaluating the $l^0$ integral by closing the contour in the lower half of the complex energy plane, the contribution arises from the Kaon pole
$z_A:\, l^0=\omega_1-i\epsilon$ with the infinitesimal $\epsilon$.
The corresponding residue reads
\begin{widetext}
\begin{eqnarray}
&&2\pi i\,\mathrm{Res}\left[I(z_A)\right]
=
\int \frac{d^{3}l}{(2\pi)^3}\,
\frac{g_{\rm eff}}{8\,m_K^3}\,
\frac{4\vec{l}^2}{P_1^0-\omega_1-\omega_2}\frac{u(P_2)\,\bar{u}(P_4)}{P_3^0-\omega_1-\omega_3} \frac{1}{2m_{\Lambda}\left(P_1^0+P_2^0-\omega_1 -\omega_{\Lambda}\right)}
\nonumber\\
&&+
\int \frac{d^{3}l}{(2\pi)^3}\,
\frac{g_{\rm eff}}{8\,m_K^3}\,
\frac{\vec{P}_1^{\,2}}
{P_1^0-\omega_1-\omega_2}\,
\frac{u(P_2)\,\bar{u}(P_4)}{2m_{\Lambda}\left(P_1^0+P_2^0-\omega_1 -\omega_{\Lambda}\right)},
\label{eq:ResLoop}
\end{eqnarray}
\end{widetext}
where $P_1^0-\omega_1-\omega_2<0$ and $P_3^0-\omega_1-\omega_3<0$, which correspond to the $K\bar K$
pair forms a bound state associated with the $\phi$ meson with the unphysical Kaon mass.
Consequently, the loop function remains real below the $NK\bar K$ threshold.

The second term on the right-hand side of
Eq.~(\ref{eq:ResLoop}) is proportional to $\vec{P}_1^{\,2}$ and vanishes
in the limit $\vec{P}_1^{\,2}\to 0$.
Consequently, only the first term contributes to the leading-order S-wave potential,
which is convergent and reads
\begin{eqnarray}
V_{LO}^s
&=&
\int \frac{d^{3}l}{(2\pi)^3}\,
\frac{g_{\rm eff}}{8\,m_K^3}\,
\frac{4\vec{l}^2}{(P_1^0-\omega_1-\omega_2)^2}\nonumber\\
&&\times \frac{u(P_2)\,\bar{u}(P_4)}{2m_{\Lambda}\left(P_1^0+P_2^0-\omega_1 -\omega_{\Lambda}\right)},\nonumber\\
&=&\int \frac{d^{3}l}{(2\pi)^3}\,
\frac{g_{\rm eff}}{8\,m_K^3}\,
\frac{4\vec{l}^2u(P_2)\,\bar{u}(P_4)}{\left(\delta -\frac{\vec{l}^2}{m_K}\right)^2} \frac{1/2m_{\Lambda}}{\Delta E-\frac{\vec{l}^2}{2m_{\Lambda}}}
\label{eq:Vloop}
\end{eqnarray}
where $\delta = m_\phi-2m_K$ and $\Delta E=m_{\phi}+m_{p}-\omega_1-m_{\Lambda}$ with unphysical masses in the Table. \ref{tab:masses}.

\textit{Unphysical case:} With the definitions $a_0=\sqrt{-m_K \delta}$ and $b^2=-2m_{\Lambda} \Delta E$, the momentum integral in Eq.~(\ref{eq:Vloop}) is evaluated using a parameter $\Lambda$, yielding
\begin{eqnarray}
   \frac{V^s_{LO} }{\mathcal{C}}&=& \arctan\left( \frac{\Lambda}{a_0}\right) \left[ \frac{a_0}{2(b^2-a_0^2)} +\frac{a_0(a_0^2-2b^2)}{(a_0^2-b^2)^2}\right]\nonumber\\
   &&+ \frac{a_0}{2(b^2-a_0^2)}\frac{a_0\Lambda}{a_0^2+\Lambda^2}\nonumber\\
   &&+ \arctan \left( \frac{\Lambda}{b}\right)\frac{b^3}{(b^2-a_0^2)^2},\label{eq:Vunphy}
\end{eqnarray}
where $\mathcal{C}=-2m_K^2g_{\rm eff}/\pi^2$. 
It is worth noting that the momentum integral in Eq.~(\ref{eq:Vloop}) is strictly UV convergent by simple power counting. Therefore, the parameter $\Lambda$  introduced here is not a mathematical regularizer for a divergence, but rather a physical scale characterizing the finite size of the interacting hadrons.
The analytical expressions reveal that the potential $V_{LO}^s$ is significantly enhanced as $|a_0|$ and $|b|$ vanish, corresponding to the $K\bar{K}$ threshold approaching the $\phi$ mass and the $\Lambda(1405)$ state becoming on-shell. 
This behavior is a signature of the kinematical enhancement inherent in the pole-promoted triangle-like mechanism.


In the limit where $|b|$ acts as a hard scale relative to $|a_0|$, the interaction simplifies to
\begin{eqnarray}
    \frac{V^s_{LO} }{\mathcal{C}}&\simeq& \frac{-3a_0}{2b^2}\arctan\left( \frac{\Lambda}{a_0}\right) + \frac{1}{b}\arctan \left( \frac{\Lambda}{b}\right) \nonumber\\
   &&+ \frac{a_0}{2b^2}\frac{a_0\Lambda}{a_0^2+\Lambda^2},
\end{eqnarray}
which is proportional to $a$ within a narrow energy window.
For the hierarchy $\Lambda\gg|b|\gg |a|$, the interaction reads
\begin{eqnarray}
     \frac{V^s_{LO} }{\mathcal{C}}
   &\simeq& \frac{\pi}{2}\left( \frac{2b-3a_0}{2b^2}\right), \label{eq:Vunphy1}
\end{eqnarray}
implying that the $\Lambda(1405)$ plays a crucial role in $N\phi$ scattering.
This behavior differs from that characteristic of long-range interactions.
In particular, the QCD van der Waals force arising from multi-gluon exchange exhibits a much steeper dependence, $V \propto p_{N\phi}^{6}$ or $p_{N\phi}^{7}$ \cite{Appelquist:1978rt,Brambilla:2017ffe}.
The derived interaction is also distinct from one-boson-exchange models or Yukawa-type interactions \cite{Kim:2021adl} and can be further tested in three-body systems \cite{Wen:2025wit}.

\begin{table}
		\caption{Inputs of the isospin-averaged hadron masses \cite{Lyu:2022imf}}
\begin{tabular}{ccc}
	\hline \hline Hadron & Lattice [MeV] & Expt. [MeV] \\
	$K$ & $524.7(2)$ & 495.6 \\
	$\phi$ & $1048.0(4)$ & 1019.5 \\
	$N$ & $954.0(2.9)$ & 938.9 \\
	\hline \hline
\end{tabular}\label{tab:masses}
\end{table}

\textit{Physical case:} For physical hadron masses, the ratio $\sqrt{m_K|\tilde{\delta}|}/m_K \simeq 0.24$ indicates that the intermediate Kaons in the triangle diagram remain non-relativistic. This validates the expansion in Eq.~(\ref{eq:Vloop}) and motivates the evaluation of the $N\phi$ scattering length using physical parameters. In this regime, $\delta$ becomes positive, introducing a pole in the propagator at momentum $k_0=\sqrt{2 m_K \delta}$. Consequently, the interaction takes the form
\begin{eqnarray}
    \frac{V_{LO}^s}{\mathcal{C}}&=& \frac{k_0}{4\left( b^2+k_0^2\right)}\left( \log \frac{\Lambda+k_0}{\Lambda-k_0} - i\pi-\frac{2k_0\Lambda}{\Lambda^2-k_0^2}\right)\nonumber\\
    &&+\frac{k_0 \left( k_0^2 +2b^2\right)}{2\left( b^2+k_0^2\right)^2}\left(\log \frac{\Lambda-k_0}{\Lambda+k_0} +i \pi\right)\nonumber\\
    &&+\frac{b^3}{\left( b^2+k_0^2\right)^2}\arctan\left( \frac{\Lambda}{b}\right),\label{eq:Vphy}
\end{eqnarray}
where, in the large-$\Lambda$ limit, the real part scales as $1/b$. This behavior aligns with the result for unphysical hadron masses in Eq.~(\ref{eq:Vunphy}), confirming that the physical $N\phi$ scattering driven by the triangle mechanism is distinct from the QCD van der Waals force. The imaginary part, arising from the $\pm i\pi$ terms, reflects the open-channel effect of the $\phi \to K\bar{K}$ decay characterized by $k_0$. Furthermore, the $\Lambda$ dependence is logarithmic and remains mild for $\Lambda \gg k_0$.

\section{Results and discussion}\label{Sect3}

According to the HAL QCD results obtained with an unphysical Kaon mass
$m_K=524.7~\mathrm{MeV}$, as listed in Table.~\ref{tab:masses}, the same Kaon mass
is adopted in the present analysis for consistency.
The $K\bar K$ threshold lies slightly above the $\phi$ mass,
leading to
\[
\delta \equiv m_\phi-2m_K = -\,\tilde\delta = -1.4(8)~\mathrm{MeV}.
\]
This small negative value indicates that the $\phi$ is marginally bound with
respect to the $K\bar K$ threshold at the unphysical mass point, which enhances
the near-threshold sensitivity of the triangle diagram.

The use of an unphysical Kaon mass also induces a corresponding modification of
the $\phi K\bar K$ coupling constant $g$.
Invoking $SU(3)$ flavor symmetry, this coupling is related to the $\rho\pi\pi$
coupling, $g_{\rho\pi\pi}$, governing the decay $\rho\to\pi\pi$.
The pion-mass dependence of this coupling can be parametrized as
$
g_{\rho\pi\pi} = \tilde c_0 + \tilde c_1\, m_\pi^2 .
$
Lattice and phenomenological studies show that $g_{\rho\pi\pi}$ varies only
mildly, at the level of $2\%\!-\!4\%$, when the pion mass is varied from
$140$ to $200~\mathrm{MeV}$~\cite{Wang:2025hew}, in agreement with earlier
analyses indicating an approximate pion-mass independence of this
coupling~\cite{Hanhart:2008mx,Guo:2011gc}.
Consequently, the modification of the $\phi K\bar K$ coupling induced by the
unphysical light-quark masses is expected to be numerically small and does not
affect the qualitative features of the near-threshold $N\phi$ interaction
discussed below.

\begin{figure}[ht]
    \centering
\includegraphics[width=0.9\linewidth]{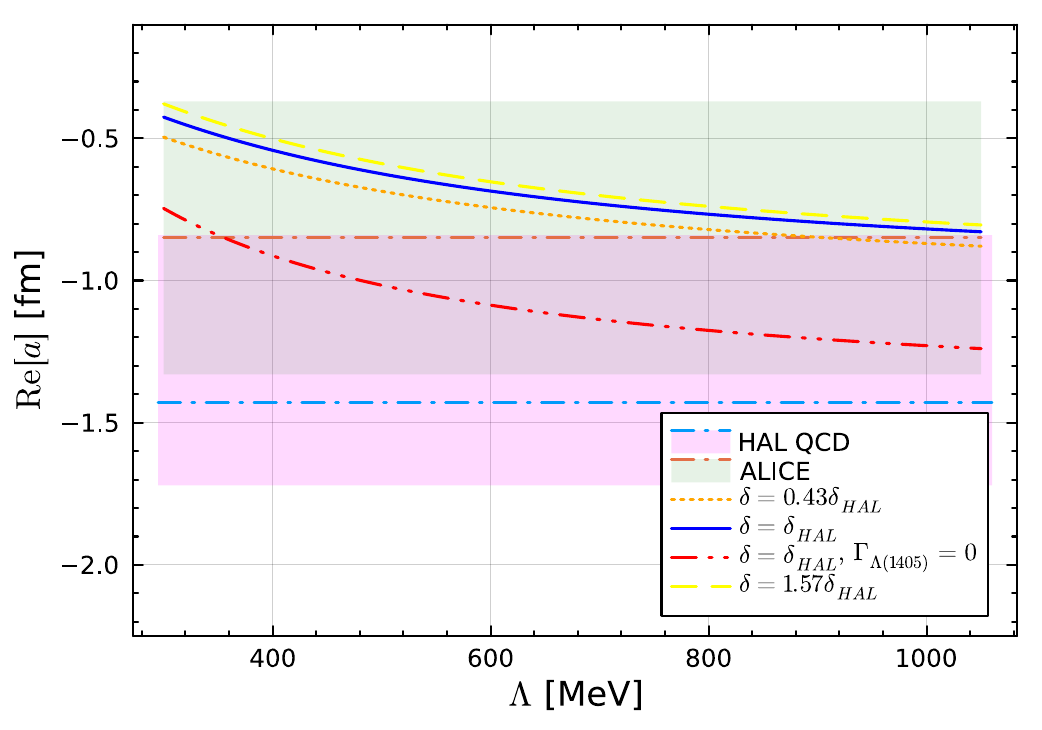}
    \caption{The $N\phi$ scattering length from two-Kaon-Exchange with $\tilde{\delta}=\left(1.4\pm 0.8 \right)\, \rm{MeV}$,  where the solid, dashed, and dotted lines correspond to the central value, lower and upper limits of $\delta$, respectively. The magenta and green bands represent the real part of the scattering lengths from HAL QCD \cite{Lyu:2022imf} and ALICE collaboration \cite{ALICE:2021cpv}, respectively.}
    \label{fig:TwoKExCh}
\end{figure}

The numerical results for the $N\phi$ scattering length are obtained by varying
the binding energy of the intermediate $K\bar K$ system to the $\phi$ meson, where the center values of $\sqrt{s}_R$ and $g_i$ ~\cite{Guo:2012vv} are adopted.
The corresponding results are shown in Fig.~\ref{fig:TwoKExCh}, where the dash-dot-dot line represents the case of switching off the imaginary part of $\sqrt{s}_R$.
It is easy to see that the difference between the dashed and dotted lines is small, which implies that the changes in $\delta$ affect the interaction in Eq. (\ref{eq:Vunphy}) only mildly. On the other hand, the obvious difference between the solid and dash-dot-dot lines is caused by keeping the imaginary part of $\sqrt{s}_R$ or not, which supports that $\Lambda(1405)$ is crucial to yield the interaction in the triangle diagram. Additionally, the calculated curves, with $\Lambda >700 \, \rm{MeV}$, show good agreement with the magenta band extracted
in the spin-$3/2$ channel at an unphysical pion mass in Ref.~\cite{Lyu:2022imf},
as well as with the green band corresponding to the spin-averaged scattering
length measured by the ALICE collaboration~\cite{ALICE:2021cpv}.

Subsequently, the leading-order S-wave potential induced by the triangle diagram
is estimated analytically in Eq. (\ref{eq:Vunphy}),
which is a scalar quantity and provides a large contribution to the
$N\phi$ scattering length.
The enhancement of the triangle
amplitude induced by the near-threshold $\Lambda(1405)$ pole is determined by the pole position $
\sqrt{s}_R$ and the coupling $g_i$
in Ref.~\cite{Guo:2012vv}.

\begin{figure}[ht]	\centering	\includegraphics[width=0.9\linewidth]{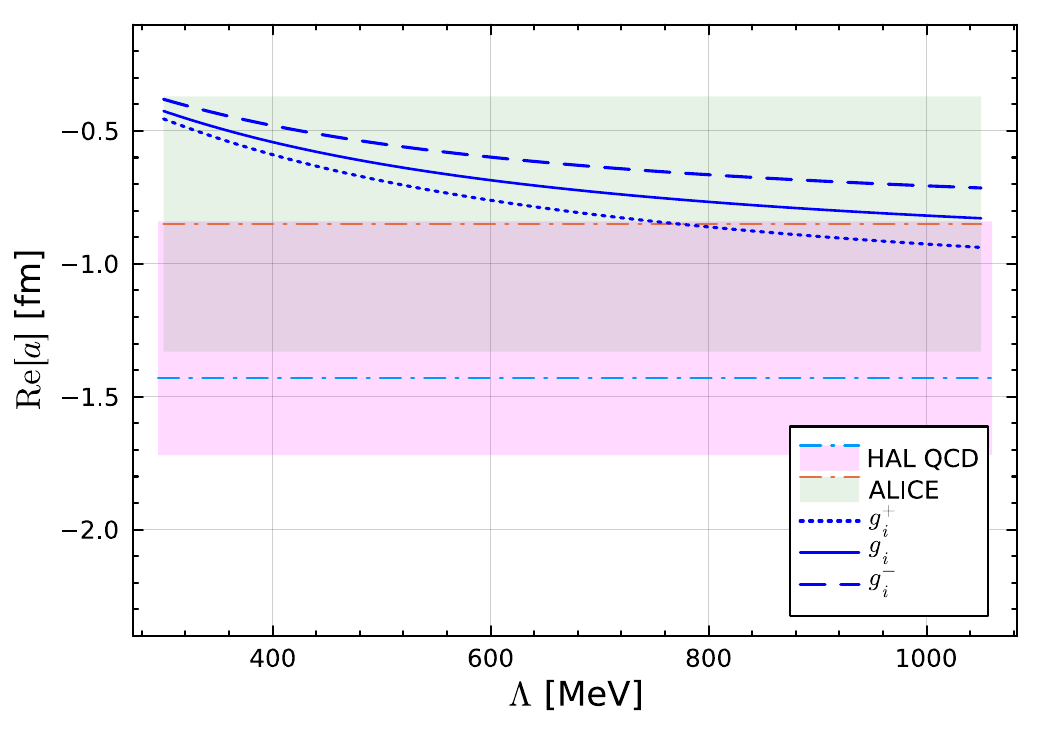}
	\caption{The $N\phi$ scattering length from two-Kaon-Exchange with uncertainty in $\Lambda(1405)$ pole. 
    }
	\label{fig:TwoKExChErr00}
\end{figure}

The uncertainty associated with $\sqrt{s}_R$ and $g_i$,
which is determined from the physical parameters and pole couplings,
propagates into the prediction of the scattering length.
This effect is illustrated in Fig.~\ref{fig:TwoKExChErr00}, where the results are
presented for the central value of $\tilde{\delta}$, and the lines
reflect the uncertainty of the $\Lambda(1405)$ pole. The discrepancy between the theoretical predictions is very small and is larger than the cases in Fig. \ref{fig:TwoKExCh}, which indicates that the $\Lambda(1405)$ dominates over the $\phi$ coupling to $K\bar{K}$ in the triangle diagram. In the meantime,
the uncertainty bands fully cover the theoretical curves.

Concerning the use of physical couplings in $V^{s}_{LO}$, the possible
modification of the coupling $g_i$ induced by the unphysical hadron masses
is estimated at the order-of-magnitude level by a mass extrapolation,
\begin{eqnarray}
\frac{g_i^{\rm unphy.\,2}}{g_i^{2}}
\simeq
\left(\frac{m_N^{\rm unphy.}}{m_N}\right)^2
\left(\frac{m_K^{\rm unphy.}}{m_K}\right)^2
\approx 1.19, \label{eq:gMass}
\end{eqnarray}
which serves only as an indicative estimate.
A quantitative determination of this dependence would require a dedicated lattice
QCD input is beyond the scope of this study.

\begin{figure}[ht]	\centering	\includegraphics[width=0.9\linewidth]{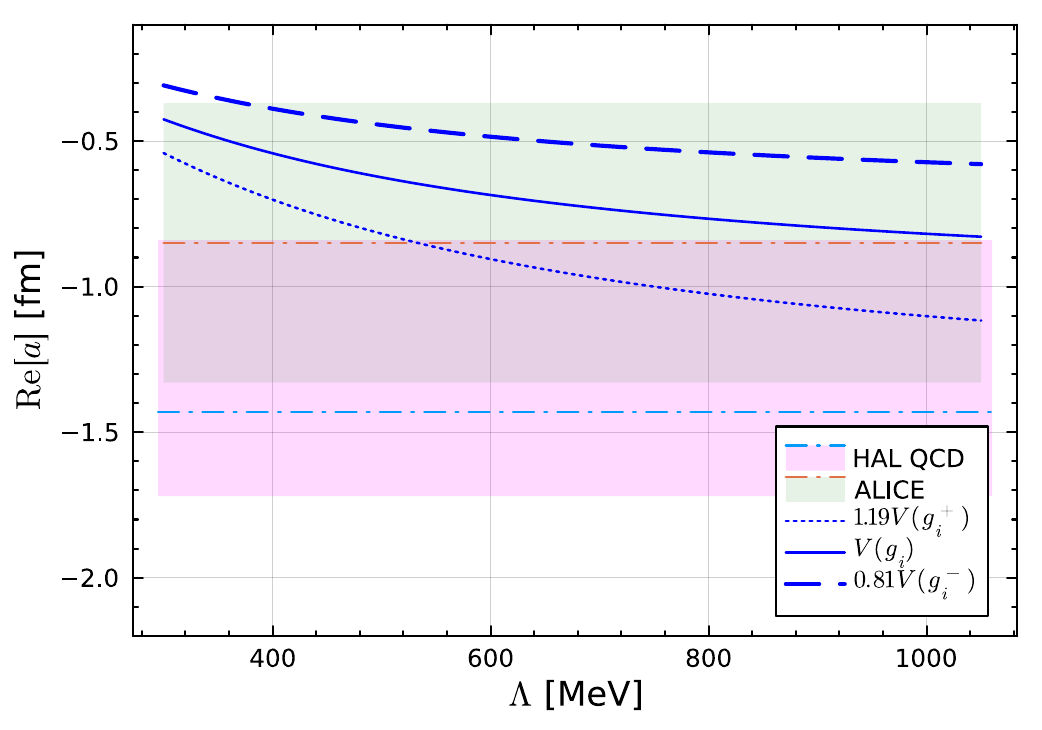}
\caption{The $N\phi$ scattering length from two-Kaon-Exchange with uncertainty in $g_i$ with unphysical hadron masses. 
    }
	\label{fig:TwoKExChErr0}
\end{figure}

The uncertainty in $g_i$ corresponds to an overall variation of the
potential at the level of $\pm 19\%$.
As a consequence, the uncertainties shown in Fig.~\ref{fig:TwoKExChErr0}
are comparable to those in Fig.~\ref{fig:TwoKExChErr00}, with the dashed and dotted
curves remaining within the corresponding error bands.
This stability indicates that the triangle-diagram–driven dynamics persists
once the quark-mass dependence of the couplings is considered.
Therefore, the conclusion on the dominance of the
two-Kaon–exchange triangle mechanism remains unaffected by reasonable
variations of the coupling constants.

In addition, the possible modification of the pion decay constant $F_\pi$ associated with the unphysical Kaon mass is taken into account.
Following Ref.~\cite{Ananthanarayan:2017yhz}, it is parameterized as
\begin{eqnarray}
F_\pi^{\rm unphy.}
=
F_\pi^{\rm phy.}
\left(1+r_{F_\pi}\,\Delta m^2\right),
\label{eq:FpiErr}
\end{eqnarray}
with $\Delta m^2 = m_K^{\rm phys.\,2}-m_K^2$ and
$r_{F_\pi}=0.53\,\mathrm{GeV^{-2}}$, as extracted from the $BE14extract$ shown in
Fig.~4 of Ref.~\cite{Ananthanarayan:2017yhz}.
Around the physical Kaon mass, the $free-fit$ results show a smooth behavior,
leading to a correction of less than $1\%$ in $F_{\pi}$.
Such a small effect is therefore treated as a higher-order perturbative
correction.
The associated uncertainty is estimated through Eq.~(\ref{eq:FpiErr}) as
\begin{eqnarray}
\left(1+r_{F_\pi}\,\Delta m^2\right)^{-n_0}
=
1.00\,({-0.03})\,({-0.06}),
\end{eqnarray}
where $n_0=2$ and $4$ correspond to the cases in which the coupling $g$
is taken to be independent of, or dependent on, $F_\pi$, respectively. This uncertainty is much smaller than the one in Eq. (\ref{eq:gMass}) and counted perturbatively.

These results suggest that the threshold dynamics driven by the
two-Kaon–exchange triangle diagram, promoted by the near-threshold
$\Lambda(1405)$ pole in the $N\bar K$ scattering, provides a natural and consistent explanation of the observed $N\phi$ scattering length.

\begin{figure}[ht]
	\centering
	\includegraphics[width=0.9\linewidth]{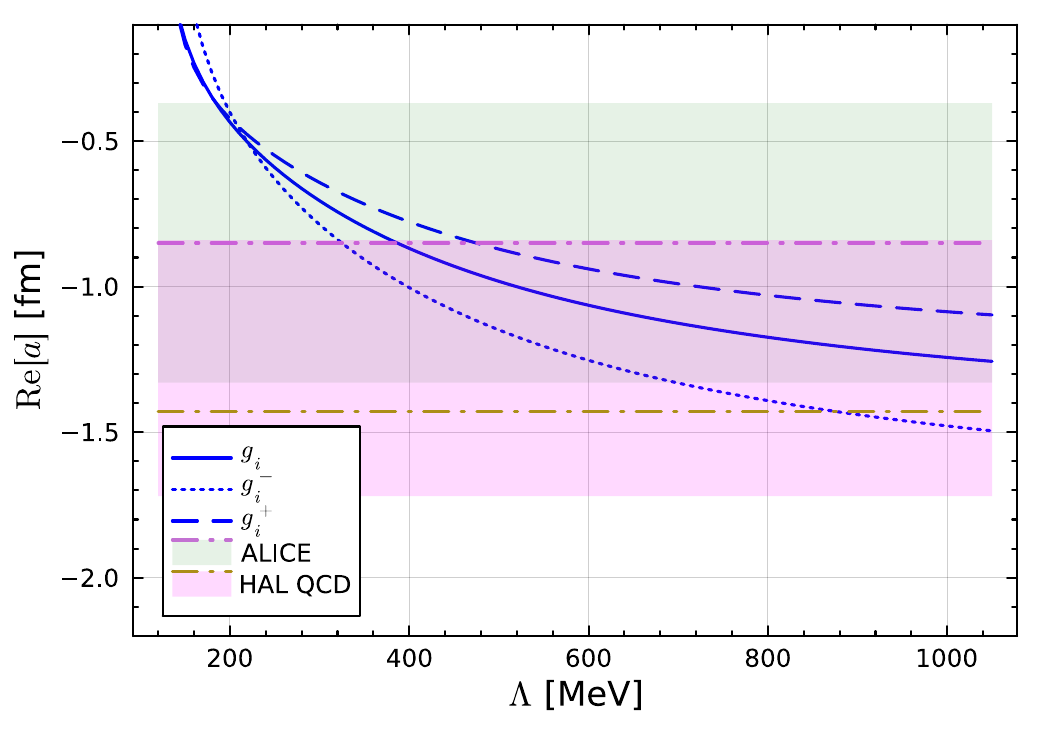}
	\caption{The real part of the $N\phi$ scattering length from Two-Kaon-Exchange with physical masses,  where the solid, dashed, and dotted lines correspond to the central value, lower and upper limits of $g_i$ and the corresponding $\sqrt{s}_R$, respectively. 
    }
	\label{fig:TwoKExChPhy}
\end{figure}

On the other hand, for physical hadron masses, the Kaons propagating in the triangle diagram remain non-relativistic.
This provides a self-consistency check for the expansion in
Eq.~(\ref{eq:Vloop}) and motivates an evaluation of the $N\phi$ scattering length
using physical masses in Eq. (\ref{eq:Vphy}).
The real part of the  scattering length is shown in
Fig.~\ref{fig:TwoKExChPhy}, where the momentum carried by the outgoing Kaon is
$|\vec{k}_K|=118.17\,\mathrm{MeV}$.

Kinematical consistency requires the parameter $\Lambda$ in the loop integral to be
larger than the typical momentum flowing through the Kaon lines.
For a small value, $\Lambda=120\,\mathrm{MeV}$, noticeable deviations from the uncertainty bands are observed.
In contrast, for $\Lambda=400$-$1000\,\mathrm{MeV}$ the real part of the
$N\phi$ scattering length remains stable and overlaps with the bands, supporting
the relevance of the triangle mechanism in the physical-mass case.

\begin{figure}[ht]
	\centering
	\includegraphics[width=0.9\linewidth]{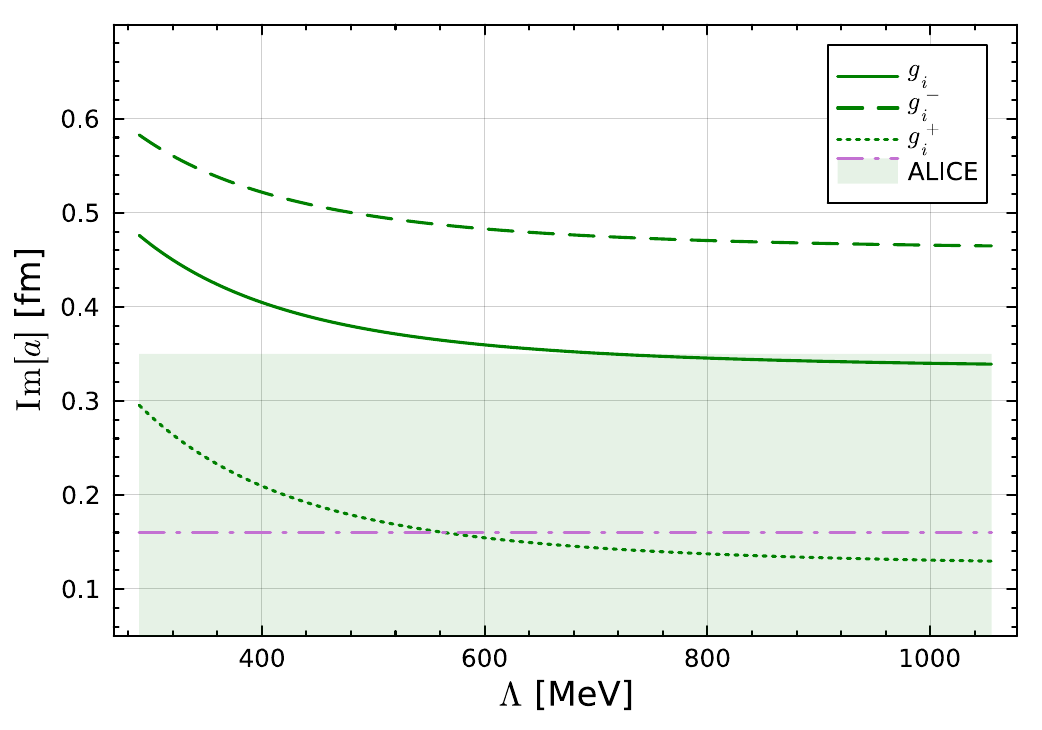}
	\caption{The imaginary part of the $N\phi$ scattering length from Two-Kaon-Exchange with physical masses,  where the solid, dashed, and dotted lines correspond to the central value, lower and upper limits of $g_i$ and the corresponding $\sqrt{s}_R$, respectively. 
    }
	\label{fig:TwoKExChPhyIm}
\end{figure}

In Fig. \ref{fig:TwoKExChPhyIm}, the imaginary part of $N\phi$ scattering length determined by the open channel effect, should be free of the $\Lambda$ in the triangle. However, the open channel effect is not evaluated directly here and approximated by the $\Lambda(1405)$ pole. When $\Lambda$ is much larger than $\vert b\vert$, the green lines show little dependence of $\Lambda$ and the dotted line overlaps with the one extracted by ALICE \cite{ALICE:2021cpv}, which indicates that the triangle diagram is non-negligible in $N\phi$ scattering.

The scattering lengths are consistent with the real parts obtained in
spin-resolved analyses, including the spin-$1/2$ result of
Ref.~\cite{Chizzali:2022pjd}, the spin-$3/2$ results of
Refs.~\cite{Lyu:2022imf,Kuros:2024dhc,Abreu:2024qqo}, as well as the spin-averaged
extraction by the ALICE collaboration~\cite{ALICE:2021cpv}.
They differ from estimates that include strong coupled-channel effects in the
correlation-function approach~\cite{Feijoo:2024bvn}.
Clarifying the origin of these differences among various model descriptions
\cite{Chizzali:2022pjd,Kuros:2024dhc,Abreu:2024qqo} will require further
experimental input and refined analyses.

The effectiveness of the triangle-diagram mechanism in $N\phi$ scattering can be
traced back to a set of well-defined dynamical conditions.
First, there is no interaction channel dominated by a conventional
one-particle-exchange contribution at low energies.
Second, the two-pion–exchange contribution in the two-point loop is
suppressed, owing to the vanishing of the leading-order
$\phi\pi\to\phi\pi$ Weinberg-Tomozawa interaction and weak interactions from other diagrams (see Appendix).
Third, the intermediate particles in the triangle diagram propagate close to
their mass shells, and the presence of a near-threshold pole in the
$N\bar K$ subsystem further enhances the amplitude.
Together, these features single out the two-Kaon–exchange triangle diagram as
the dominant low-energy mechanism.

Such threshold-enhanced triangle dynamics is not unique to the $N\phi$ system
and has been widely discussed in heavy exotic hadrons.
Representative examples include the role of $D\bar D$ rescattering in
$D\bar D\pi$ loops for the $X(3872)$
\cite{Dai:2019hrf,Yan:2022eiy,Cao:2024nxm},
the contribution of $\Sigma_c\bar D\pi$ loops to positive-parity
$P_c(4457)$ states \cite{Peng:2020gwk,Wu:2024bvl},
and the importance of $DD\pi$ loops in the description of the $T_{cc}$ state
\cite{Fleming:2021wmk,Yan:2021wdl,Jia:2022qwr,Jia:2023hvc,Dai:2023mxm}.
In these systems, the proximity of two- and three-body thresholds leads to
non-perturbative enhancements analogous to those identified here.

A similar mechanism is expected to operate in channels where
$D\bar D\pi$ couples to $Z_c(4020)$ or $X(4020)$, promoted by the nearby
$X(3872)$ or $W_{c1}(3872)$ poles
\cite{Peng:2023lfw,Ji:2025hjw}.
In such cases, the triangle diagram generates an interaction that
plays a role comparable to that of scalar-meson exchange, supplementing the
axial-vector exchange mechanism \cite{Yan:2021tcp}.
Related threshold effects have also been discussed for hidden-charm and
hidden-strangeness pentaquark states, where the $P_c(4457)$ and $P_{cs}(4560)$
states are enhanced by lower-lying $P_c(4312)$ and $P_{cs}(4338)$ poles through
$\bar D\Sigma_c\pi$ and $\bar D\Xi_c\pi$ loops, respectively
\cite{Peng:2022iez,Du:2019pij,Fernandez-Ramirez:2019koa,Wang:2019hyc,Liu:2019tjn,
Wang:2019nvm,Chen:2020uif,Yan:2021nio,Xiao:2021rgp,Yan:2022wuz,Nakamura:2022gtu,Zhu:2022wpi,Yang:2022ezl}.

Finally, in open-charm pentaquark systems, near-threshold poles in the $ND$ and
$N\bar D$ channels can further promote the formation of $ND^\ast$ and
$N\bar D^\ast$ states through analogous triangle mechanisms
\cite{Peng:2021hkr,Yan:2023ttx,Qiao:2024acm}.
These examples collectively illustrate that the triangle-diagram mechanism
identified in this study represents a generic manifestation of
threshold-driven three-body dynamics rather than a system-specific anomaly.
The $N\phi$ system thus provides a particularly transparent realization of
threshold-enhanced triangle dynamics.

In the dibaryon sector, a closely related dynamical mechanism has been invoked
to interpret the $d(2380)$ as a $\Delta\Delta$ molecular state, where the existence of a near-threshold $N\Delta$ bound state plays a crucial role
\cite{Gal:2013dca}.
A similar picture also emerges in the $IJ=21$ and $12$ $N\Delta$ scattering
channels when poles in the $NN$ subsystem are taken into account
\cite{Gal:2014zia}.
However, the situation for the $d(2380)$ is considerably more involved: the resonance is located between the $N\Delta\pi$ and $\Delta\Delta$
thresholds, and the large intrinsic width of the $\Delta$ renders a clean separation between two-body and three-body dynamics highly nontrivial
\cite{Peng:2021hkr}.
This contrasts with the $N\phi$ system, where the thresholds are much better
separated, and the triangle mechanism can be isolated more transparently.

Another instructive comparison is provided by the $p\Omega$ system, for which a bound state has been reported by the HAL QCD collaboration
\cite{HALQCD:2014okw,HALQCD:2018qyu}.
When the $\Omega$ couples to the $\Xi\bar K$ channel, a triangle diagram
naturally enters the $p\Omega$ interaction, bearing qualitative similarity to the mechanism discussed for $N\phi$ scattering.
A crucial difference is that $m_{\Xi}+m_{\bar{K}}-m_{\Omega}=143\,\rm{MeV}$ leads to a relativistic Kaon in the triangle diagram and a different loop integral over $l^0$ in Eq. (\ref{eq:Il}).
As a result, the interaction differs from the one in Eq. (\ref{eq:Vphy}) and is instead tied to coupled-channel dynamics
\cite{Sekihara:2018tsb}.
This feature leads to a stronger attractive interaction in the
$p\Omega$ system than in the $N\phi$ case. In the meantime,
the comparison suggests that the attraction generated by a convergent triangle diagram alone is insufficient to form a bound state at an unphysical Kaon mass
around $500\,\mev$.

In the light-meson sector, a pronounced enhancement has been observed near the
$\omega\phi$ threshold \cite{BESIII:2012rtd,Dorofeev:2023war}.
This structure may plausibly be driven by a $K_1(1270)\bar K$ triangle loop, given that the $\omega K$ interaction vanishes at leading order in the Weinberg-Tomozawa interaction. Such a threshold mechanism provides a complementary perspective on the origin
of the enhancement and may offer new insight into the long-standing discussion
of the glueball-like nature of the $f_0(1710)$.


\section{Summary}\label{sect4}
In summary, we investigate low-energy $N\phi$ scattering via a triangle diagram promoted by the near-threshold $\Lambda(1405)$ pole, in the absence of direct spin interactions. The scattering length, evaluated from the triangle loop integral, is found to be consistent with the lattice QCD results from the HAL QCD collaboration~\cite{Lyu:2022imf}. Furthermore, using the analytic expression derived from the triangle diagram with physical hadron masses, the obtained $N\phi$ scattering length shows a clear overlap with the experimental extraction by the ALICE collaboration~\cite{ALICE:2021cpv}.

In the low-energy region, the triangle loop integral relies on the $\Lambda(1405)$.
This behavior indicates that the interaction driven by the promoted triangle
diagram is different from both the Van der Waals force
\cite{Appelquist:1978rt,Brambilla:2017ffe} and the long-range tail of two-pion
exchange \cite{Lyu:2022imf}.

The same dynamical mechanism may also be relevant for understanding other
hadronic systems, such as $P_c(4457)$, $Z_c(4020)/X(4020)$, $d(2380)$, and the
glueball candidate $f_0(1710)$, where near-threshold poles and intermediate three-body states play an essential role.

Overall, the interaction driven by a promoted triangle diagram provides a unified framework to connect low-energy two-body scattering with underlying three-body dynamics.
This mechanism offers a complementary perspective on hadron spectroscopy in
the nonperturbative regime and can be further tested with future experimental
measurements and high-precision lattice QCD simulations.

\section*{Acknowledgement}
We would like to thank Ju-Jun Xie, Xu Zhang, and Y. Lyu
for the fruitful discussions. This research is supported by the National Natural Science Foundation of China under Grant No. 12305096, the Fundamental Research Funds for the Central Universities under Grant
No. SWU-XDJH202304 , No. SWU-KQ25016 and Chongqing Natural Science Foundation under Project No. CSTB2025NSCQ-GPX0516.

\bibliography{refs.bib}

\section*{Appendix: Two-pion and Kaon exchange in $N\phi$ scattering}

\begin{figure}[ht]	\centering	\includegraphics[width=0.9\linewidth]{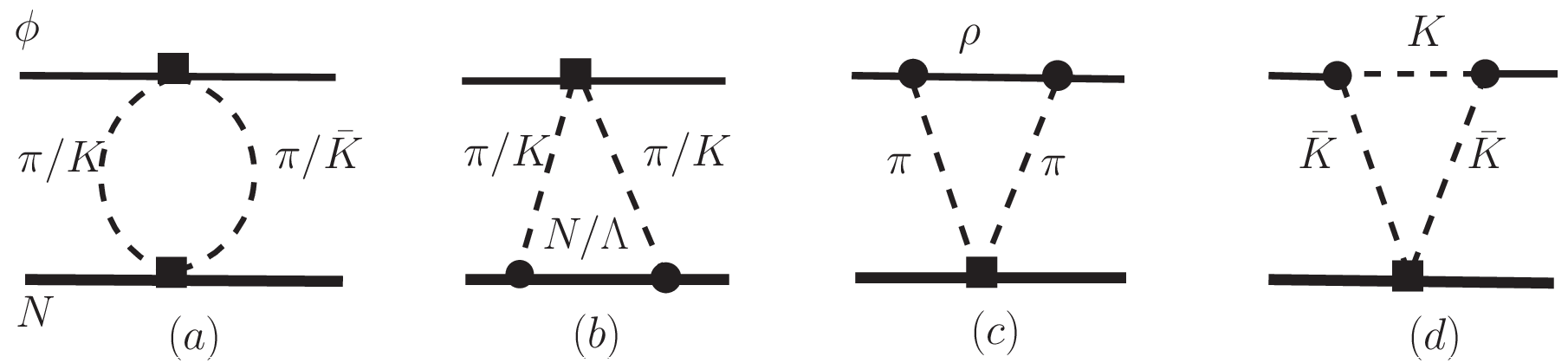}
	\caption{The Feynman diagrams of $N\phi$ scattering with two-pion and Kaon exchange. The rectangle and the solid circle stand for the S- and P-wave transitions, respectively. 
    }
	\label{fig:TwoPiandK}
\end{figure}

In this appendix, we examine possible contributions from two-pion and Kaon
exchange to the $N\phi$ interaction.
The corresponding Feynman diagrams are shown in Fig.~\ref{fig:TwoPiandK}.
Both $S$- and $P$-wave transitions are in principle allowed.
The $S$-wave transition is governed by the
Weinberg-Tomozawa (WT) term at leading order in chiral perturbation theory,
while the $P$-wave transition arises from the axial coupling.

However, the WT interactions for elastic $\phi\pi$, $\phi K$, and
$\phi\bar K$ scattering vanish at leading order
\cite{Roca:2005nm}.
As a consequence, the contribution from the two-pion exchange diagram shown in
Fig.~\ref{fig:TwoPiandK}(a) identically vanishes.
This situation is in contrast to the case of two-pion exchange in
$J/\psi N$ scattering, where a nonvanishing $J/\psi\pi \to J/\psi\pi$
transition exists, as illustrated in Fig.~1(c) of
Ref.~\cite{TarrusCastella:2018php}.
Similarly, the Kaon-exchange contribution shown in
Fig.~\ref{fig:TwoPiandK}(b) also vanishes at this order.
This again differs from the mechanism in Fig.~1(d) of
Ref.~\cite{TarrusCastella:2018php}, where the corresponding meson-hadron
transition is allowed.
Therefore, neither two-pion nor Kaon exchange generates a long-range
interaction in $N\phi$ scattering in two-point loop.
These support the conclusion drawn in the main text: the $N\phi$ interaction near threshold is not primarily a result of conventional
one- or two-meson exchange mechanisms, but is instead driven by the promoted
triangle diagram involving near-on-shell intermediate states.

In Figs.~\ref{fig:TwoPiandK}(c) and (d), the $\phi$ meson couples to
$\rho\pi$ and $K\bar K$, respectively.
According to the PDG \cite{ParticleDataGroup:2024cfk}, the branching ratio
of $\phi\to K\bar K$ is about 5.4 times larger than that of
$\phi\to \rho\pi$, despite the fact that the available phase space for
$\phi\to K\bar K$ is  smaller than that for
$\phi\to \rho\pi$.
This feature indicates a much stronger effective coupling of the $\phi$
meson to the $K\bar K$ channel compared to the $\rho\pi$ channel.

In the baryon exchange in the triangle diagrams, the poles ($N(1440),\,\Delta$) in $N\pi$ scattering are known to be broad
and located far away from the $N\pi$ threshold.
As a consequence, the enhancement induced by $N\pi$ rescattering in
$N\phi$ scattering is much weaker than that generated by
$N\bar K$ rescattering promoted by the near-threshold
$\Lambda(1405)$ pole. Meanwhile, the intermediate $N$ and $\Lambda$ in diagram~(b) are away from their mass-shell. As a result, the $\rm N^3$LO two-pion exchange in Ref.
\cite{Hatsuda:2025djd} is not kinematically and differs from the mechanism in the main text.
Therefore, the contribution from the diagram in
Fig.~\ref{fig:TwoPiandK}(c) is expected to be  suppressed compared
to that in Fig.~\ref{fig:TwoPiandK}(d).

Furthermore, the proposed mechanism is distinct from the conventional two-pion exchange in kinematics, dynamics, and its absorptive nature.
Kinematically, there is a clear separation of scales. The effective range of the two-pion exchange is governed by the hard scale $1/(2m_\pi)$. In contrast, the near-on-shell propagation of the intermediate Kaons introduces new soft dynamical scales, characterized by the momenta $|a_0|=\sqrt{-m_K \delta}$ and $|b|=\sqrt{-2m_\Lambda \Delta E}$. We focus on interactions within the range of order $1/|b|$, where the long-range physics of the triangle diagram dominates.
Dynamically, the two mechanisms differ fundamentally. The two-pion exchange is suppressed in the low-energy region due to the vanishing leading-order Weinberg-Tomozawa interaction in the $\phi\pi \to \phi\pi$ channel. Conversely, the two-Kaon exchange is significantly promoted by the non-perturbative three-body dynamics involving the near-threshold $\Lambda(1405)$ pole.
Crucially, unlike the real-valued potential often assumed in simple exchange models, the two-Kaon mechanism naturally generates an imaginary part for the $N\phi$ scattering length due to the open channel effects and the complex pole structure. As presented in the Fig. \ref{fig:TwoKExChPhyIm}, our calculated imaginary part is in good agreement with the femtoscopic extraction by the ALICE collaboration. This consistency with experimental data provides strong support for the dominance of the pole-enhanced triangle mechanism over the suppressed two-pion exchange in the near-threshold region.

\end{document}